\documentclass[aps,notitlepage,pra,onecolumn]{revtex4-1}
\usepackage{color,hyperref}

\begin{document}
\title{Perspective: Quantum Thermodynamics}
\date{\today}

\author{James Millen}
\affiliation{Faculty of Physics, University of Vienna, VCQ, Boltzmanngasse 5, 1090 Vienna, Austria}
\author{Andr\'e Xuereb}
\affiliation{Department of Physics, University of Malta, Msida MSD 2080, Malta}
\affiliation{Centre for Theoretical Atomic, Molecular and Optical Physics, School of Mathematics and Physics, Queen's University Belfast, Belfast BT7\,1NN, United Kingdom}

\begin{abstract}
Classical thermodynamics is unrivalled in its range of applications and relevance to everyday life. It enables a description of complex systems, made up of microscopic particles, in terms of a small number of macroscopic quantities, such as work and entropy. As systems get ever smaller, fluctuations of these quantities become increasingly relevant, prompting the development of stochastic thermodynamics. Recently we have seen a surge of interest in exploring the quantum regime, where the origin of fluctuations is quantum rather than thermal. Many questions, such as the role of entanglement and the emergence of thermalisation, lie wide open. Answering these questions may lead to the development of quantum heat engines and refrigerators, as well as to vitally needed simple descriptions of quantum many-body systems.
\end{abstract}
\maketitle

\section{Introduction}
Thermodynamics presents us with an effective picture of processes occurring in complex systems, describing the bulk properties of the system without being concerned with its microscopic details. Quantities such as the temperature of a system, the amount of work that can be extracted from it, or the heat it dissipates, reduce the description of systems consisting of untold numbers of particles to a handful of parameters. As a consequence of this `bird's eye view,' thermodynamics is widely applicable, and its laws seem to be obeyed by every process occurring in the macroscopic world.

The downside of this macroscopic description is that thermodynamics necessarily deals with average quantities. Whilst being a valid approach when the system at hand is composed of a macroscopic number of particles, it starts losing accuracy as the system size decreases and fluctuations around these average quantities, due to thermal motion, become relevant. Stochastic thermodynamics picks up where the macroscopic description starts to fail, and gives a deeper insight into the fluctuations of thermodynamic quantities. It also moves beyond the equilibrium situations associated with thermodynamics, and can describe the behaviour of systems that are held out of equilibrium~\cite{seifert_review}. These considerations are vital if considering nanoscale or biological machines.

However, when dealing with even smaller systems, quantum effects come into play; fluctuations are no longer just thermal in their origin but quantum. In this regime several questions emerge; it is not clear why the time-reversible, unitary dynamics that describes quantum processes should lead to a system ever reaching equilibrium, let alone why such a system will thermalise (reach a state that can be described by a few quantities such as temperature)~\cite{eisert_review}. Furthermore, the link between classical thermodynamic and information-theoretic quantities like entropy suggests that quantum phenomena such as entanglement could play an important role in quantum thermodynamics, a role which is not yet fully understood. Indeed, it is a challenge to even define and measure thermodynamic quantities for microscopic, quantum systems \cite{mehboudi15, viisanen15, rossnagel15, chiara15}.

Upon moving from the macroscopic, classical, world to the microscopic, quantum, realm, it is natural to ask whether the laws of thermodynamics retain their place. The \emph{zeroth law} of classical thermodynamics states that if two systems are thermalised with a third, then all three are thermalised with each other, which in the quantum regime is translated to the statement that, given a closed system governed by a particular Hamiltonian, there exists a family of states parametrised by temperature (thermal states) from which it is not possible to extract work~\cite{brandao15}.
The \emph{first law} is a statement of conservation of energy, i.e., the change in internal energy of a system during a process is equal to the heat supplied minus the amount of work done by the system. In the theory of quantum thermodynamics this can be used to define the allowed thermodynamic operations on closed quantum systems as energy-conserving unitary operations~\cite{brandao15}.
The \emph{second law} can be cast in several ways, e.g., that entropy increases when undergoing an irreversible process, that heat cannot flow from a cold bath to a hot bath, or that free energy can only decrease. This last formulation was adapted to the quantum setting, multiplying into a whole family of second laws~\cite{brandao15} restricting which thermodynamic processes can take place.
The \emph{third law} can be also stated in several ways~\cite{hanggi06}, e.g., it is impossible to reduce the temperature of a system to zero in a finite time, and sheds light on the rate at which thermodynamic processes happen~\cite{masanes14}. There is active debate as to whether it is possible to violate the third law in quantum systems~\cite{cleuren12, levy12, levy12b, kolar12}.
This discussion considers only closed quantum systems. Studies of the thermodynamics of open quantum systems are much more recent; see, e.g., Ref.~\cite{esposito15,gasparinetti14,schaller14,gaspard15,carrega15,xuereb15}.

In this article we aim to highlight some of the recent key results and open problems in the rapidly-evolving field of Quantum Thermodynamics, with particular reference to the recent Focus Issue on Quantum Thermodynamics in New Journal of Physics. Complementary points of view can be found in more technical articles reviewing the whole field~\cite{vinjanampathy15}, the role of quantum information in quantum thermodynamics~\cite{goold15a}, thermalisation in closed quantum systems~\cite{gogolin15}, and symmetry breaking in finite quantum systems~\cite{birman13}.

\section{Fluctuation theorems}
Stochastic thermodynamics~\cite{seifert_review} describes the fluctuation of thermodynamical quantities by considering individual trajectories of an evolving system. It is applicable when the fluctuations are appreciable, i.e. in small systems such as colloids or microscopic biological settings. Stochastic thermodynamics has led to the discovery of fluctuation relations~\cite{jarzynski_review}, that bound processes where systems are driven out of equilibrium. The celebrated Jarzynski~\cite{jarzynski97_1,jarzynski97_2} and Crooks~\cite{crooks99,crooks00} relations link the free energy difference between states with the work done in transforming between them, and have been experimentally verified in several systems~\cite{hummer01, liphardt02,collin05,douarche05,harris07,saira12,toyabe10, koski14_1}. These relations also hold unmodified in closed quantum systems~\cite{mukamel03,esposito_review,campisi_review}, with slight modifications when the system is open~\cite{chernyak04, cuetara15, venkatesh15}.

In the quantum setting, work is not an observable~\cite{allahverdyan05,campisi_review}, and must instead be inferred by performing projective~\cite{huber08} or interferometric~\cite{dorner13,mazzola13} measurements, or by measuring optical spectra~\cite{heyl12}. Nonetheless, quantum work fluctuations have been measured in molecular~\cite{batalhao14} and trapped-ion~\cite{an15} systems, with proposals utilising superconducting circuits~\cite{brito15}, and exchange fluctuations have been measured in electronic systems~\cite{utsumi10,nakamura11}. Interferometric techniques have also been extended to measuring the heat exchange occuring in a quantum process~\cite{goold14}. Open questions include non-linear quantum fluctuations~\cite{huber08,campisi_review}, exploiting quantum information to produce work~\cite{maruyama_review}, the use of feedback in quantum systems~\cite{morikuni11,brandner15,funo15}, and the potential application of fluctuation relations to quantum computing~\cite{ohzeki10, spilla15}. 

\section{The role of quantum information}
The second law of classical thermodynamics distinguishes between reversible processes, which do not change the entropy of a system, and irreversible ones, where work done is dissipated as heat, increasing the entropy of the system. It has been pointed out~\cite{horodecki02,brandao08,lostaglio15_1} that the role played by entropy in classical processes is analogous to that of entanglement in quantum processes: the relative entropy of entanglement, a measure of distinguishability between two states, must increase during a thermodynamic process. There are some subtleties in the issue of whether the free energy is a useful quantity to consider in the presence of coherences~\cite{horodecki13,lostaglio15_2}. However, the maximum extractable \emph{averaged work} is equal to the change in free energy for quantum systems~\cite{skrzypczyk14}, an identical result to its classical analogue. Entanglement is a necessary by-product of generating work from an ensemble of thermal states~\cite{alicki13, huber15, bruschi15}. However, in direct analogy with the classical Carnot engine, it is possible to extract maximal work without generating entanglement at the expense of power~\cite{hovhannisyan13}.

Information plays a vital role in both classical~\cite{parrondo15} and quantum~\cite{goold15a} thermodynamics; the famous Maxwell's Demon and the Szilard Engine~\cite{maruyama_review} seem to show that knowledge about a system allows one to extract work from the system, seemingly endlessly and without an increase in entropy, apparently violating the second law. These systems have been realised experimentally with colloidal particles~\cite{toyabe10,roldan14} and single electrons~\cite{koski14_1,koski14_2}, demonstrating work extraction and an apparent decrease in entropy. The resolution to this apparent paradox, as realised by Landauer, is that ``Information is Physical''~\cite{landauer91} and must be stored somewhere. It is in erasing such information that dissipation~\cite{landauer61,maruyama_review,parrondo15} and irreversibility creep in, restoring the second law; this is known as Landauer's Principle. This creation of heat through information erasure has been experimentally verified~\cite{berut11,orlov12,jun14}, and it has also been shown that if the information storage is entirely reversible then a vanishing amount of heat is dissipated~\cite{orlov12,snider12}.

The Maxwell's Demon and Szilard Engine thought experiments acquire a fundamentally different flavour in quantum mechanics, since measurement disturbs a quantum state. A further issue is which components of the system are to be considered quantum. A quantum demon can extract more work from a quantum system than a classical demon~\cite{lloyd97,zurek03}; its efficiency over a classical demon increases with the amount of quantum correlations present, as measured by the discord~\cite{zurek03, maruyama_review}, and is degraded by decoherence. Further issues involve the problem of inserting a partition (as in the Szilard engine) without altering the energy spectrum of the system~\cite{kim11}, and the indistinguishability of quantum particles, which affects the work one can extract~\cite{kim11}. It has also been shown that if a quantum memory is used in the operation of a Szilard engine then work can be extracted~\cite{park13}. Despite several suggestions for implementing these quantum thought experiments~\cite{strasberg13, elouard15}, there is as yet no realisation.

There was some early suggestion that Landauer's Principle may not hold for strongly-interacting quantum systems~\cite{allahverdyan00,allahverdyan01} due to system--bath entanglement~\cite{horhammer05,hilt09}; this would have serious consequences such as the potential for perpetual motion~\cite{nieuwenhuizen02}, and a host of issues in quantum information processing~\cite{hilt11}. However, it has been shown that this principle holds in both weakly-~\cite{piechocinska00} and strongly-interacting~\cite{hilt11} quantum systems, even in out-of-equilibrium scenarios~\cite{goold15b}. The conversion of information to work has been measured in a quantum system, and Landauer's Principle verified at the level of individual quantum logic gates~\cite{silva14}. A version of Landauer's Principle exploiting the properties of a quantum memory allows the work cost of erasure to become \emph{negative}~\cite{delrio11}.

\section{Equilibration \& Thermalisation}
Macroscopic systems driven out of equilibrium, either through a sudden `quench' of one or more parameters, or through some other means of driving, whether periodic~\cite{kitagawa10,goldman14} or not, tend to reach an equilibrium state that depends only on the energy of the initial state; the origin of this process lies in the non-linear dynamics of systems with large numbers of particles. Quantum systems, however, are constrained by the first law to unitary, linear, and time-reversible operations~\cite{eisert_review}. They always have constants of motion, as opposed to the classical case where constants of motion are only present in integrable systems. Equilibration does tend to occur in terms of expectation values of observables or the outcomes of (generalised) measurements made on either the entire system or a subsystem, but predicting the timescale on which this happens is fraught with difficulty~\cite{eisert_review,gogolin15,goldstein15}.

Furthermore, classical systems also thermalise at equilibrium, reaching a state of maximum entropy that can be described by the number of particles and a temperature, as a consequence of the second law of thermodynamics. Thermalisation of quantum systems is described as an approach to a thermal state characterised only by the number of particles and the total energy~\cite{rigol07}. Integrable systems, such as finite one-dimensional chains of hard-core non-interacting bosons, are not predicted to thermalise in this sense, but rather to approach a Generalised Gibbs Ensemble~\cite{wouters14,pozsgay14}; similar conclusions can be drawn about non-Markovian open quantum systems~\cite{yang14}. Indeed, experiments with ultracold atoms have shown equilibration to a state with distinct multiple momenta~\cite{kinoshita06} and apparent multiple temperatures~\cite{langen15}. Integrability itself can be a difficult concept to define in the quantum setting~\cite{caux11}.

The Eigenstate Thermalization Hypothesis, which may only hold for non-integrable systems, proposes that every eigenstate of the Hamiltonian of a quantum system contains properties associated with a thermal state, which at short times after a quench are hidden by coherence, and at long times revealed through dephasing~\cite{rigol08, reimann15}. The universal applicability of this hypothesis is a topic of much debate~\cite{rigol09,steinigeweg14}, and other thermalization mechanisms are suggested~\cite{riera12}. The timescale of thermalisation in many-body systems is not well-understood. Systems may pre-thermalise~\cite{berges04}, i.e., appear to reach a metastable equilibrium state on short timescales~\cite{gring12}, with the true thermal state being reached on longer timescales~\cite{marcuzzi13}. In systems exhibiting many-body localisation, transport is strongly suppressed and thermalisation breaks down~\cite{nandkishore2015}.

\section{Quantum thermodynamic machines}
Understanding the classical laws of thermodynamics led to the development of the steam engine, i.e., a device for converting one form of energy (for example, heat) into another (work), which drove the industrial revolution. Classical heat engines exist across a wide variety of scales, from combustion engines to molecular motors~\cite{serreli07,ragazzon15}. Can an analogous development take place in the quantum regime?

The very smallest classical heat engines have been implemented using optically trapped microparticles in liquid~\cite{blickle11}, and others are proposed using nanoparticles trapped in vacuum~\cite{dechant15}. These systems highlight the roles of fluctuations; along individual trajectories, energy may flow from cold into hot heat baths---the direction of work only follows the second law on average. The fluctuating interaction between such small systems and their surrounding bath can be captured by monitoring the particles' Brownian motion. Non-equilibrium situations have been studied where the particle is hotter than the bath that surrounds it~\cite{rings10,millen14}, leading to several distinct bath temperatures in the underdamped regime~\cite{millen14}. The quantum theory of Brownian motion, based on quantum fluctuations, is distinct from classical Brownian motion~\cite{hanggi06}. It predicts that the timescales of fluctuations (noise) and dissipation (friction) are different, unlike the classical case~\cite{nieuwenhuizen02}. The consequences of this are not fully understood, and it could have a profound effect on quantum Brownian motors~\cite{hanggi09}. Thermoelectric heat engines formed from systems of quantum Hall conductors~\cite{sanchez15} or single-electron quantum dots are ideal candidates for this~\cite{kennes13,bergenfeldt14,sothmann_review,berg15,sanchez13}, as recently demonstrated~\cite{thierschmann15}, but the effect has also been observed using ultracold atoms~\cite{brantut13} and a semiconductor microcavity~\cite{klembt15}. Being able to understand the transport of heat~\cite{nejad15} as well as to convert it into useful work in microcircuits would be of great technological importance, as would be understanding the use and limitations~\cite{aberg14,ng15} of coherent or quantum catalysts, i.e., auxiliary systems used to perform work with the minimal possible disturbance to the catalyst itself.

Quantum analogues of various types of thermal machine have been studied~\cite{nori_review, malabarba15, brunelli15, uzdin15}. There are proposals for realizing quantum Otto heat engines with trapped ions~\cite{abah12}, optical~\cite{alecce15} and optomechanical systems~\cite{zhang14, elouard15}, solid-state systems \cite{campisi15}, and single molecules~\cite{hubner14}; and Nernst engines using the quantum Hall effect~\cite{sothmann14}, with observation of the Seebeck effect in an ultracold paramagnetic material~\cite{wu15}. Correlation and entanglement between the system and bath effect the work that a quantum engine can produce \cite{gallego15}, and some quantum engines are predicted to surpass their classical counterparts in terms of efficiency, and even the classical Carnot limit~\cite{rossnagel14,altintas14}. Quantum effects are also predicted to enhance the capabilities of quantum batteries~\cite{binder15} and overcome the friction~\cite{plastina14} that arises from non-adiabatic operation of realistic engines~\cite{delcampo14}. Quantum refrigerators have also been theoretically studied~\cite{kosloff14,uzdin15}, with the prediction that a single qutrit could be used to cool a qubit~\cite{linden10}---a truly tiny refrigerator. At this size scale, finite-size effects can come into play as well as quantum coherences; both these effects can conspire to fundamentally limit the amount of work that can be extracted from quantum systems~\cite{horodecki13,skrzypczyk14,gemmer15}. When operating away from Carnot efficiency, however, the presence of quantum correlations has been shown to sometimes be beneficial~\cite{brunner14} for refrigeration and transport.

\section{Conclusions and outlook}
Classical thermodynamics is extremely successful at predicting the average behaviour of large, complex systems of particles. It represents an enormous simplification over accounting for the microscopic behaviour of such systems. Stochastic and quantum thermodynamics go beyond this, the former discussing thermal fluctuations and non-equilibrium dynamics, and the latter accounting for quantum uncertainties~\cite{allahverdyan15} and correlations. We are now increasingly using quantum physics to create quantum technologies. In parallel, the miniaturisation of technology makes it vital for us to be able to understand the thermodynamics of microscopic, quantum systems. Since simulating ever larger quantum many-body systems requires an exponential increase in computational power (as compared to a linear increase for classical systems), the ongoing challenge~\cite{bhaseen15} to find a simplified thermodynamic description of complex quantum systems is more relevant than ever before.

\section*{Acknowledgements}
We are grateful for discussions with Janet Anders, L\'idia del Rio, and Mathis Friesdorf. JM would like to acknowledge support from the Marie Sk\l{}odowska-Curie Action H2020-MSCA-IF-2014. This work was partly supported by the European COST network MP1209.

\end{document}